\documentclass[runningheads]{llncs}
\usepackage{amsfonts} 
\usepackage{graphicx}
\usepackage{pbox}
\usepackage{amsmath} 
\usepackage{setspace}
\usepackage{hyperref}
\usepackage{orcidlink}
\begin{document}
\title{NISF: Neural Implicit Segmentation Functions}
\titlerunning{NISF: Neural Implicit Segmentation Functions}

\author{Nil Stolt-Ansó\inst{1,2}\orcidlink{0009-0001-4457-0967} \and
Julian McGinnis\inst{2}\orcidlink{0009-0000-2224-7600} \and 
Jiazhen Pan\inst{3}\orcidlink{0000-0002-6305-8117} \and \\
Kerstin Hammernik\inst{2,4}\orcidlink{0000-0002-2734-1409} \and 
Daniel Rueckert\inst{1,2,3,4}\orcidlink{0000-0002-5683-5889} }
%
\authorrunning{N. Stolt-Ansó et al.}
%
\institute{Munich Center for Machine Learning, Technical University Munich
\and School of Computation, Information and Technology, Technical University Munich
\and School of Medicine, Klinikum Rechts der Isar, Technical University Munich
\and Department of Computing, Imperial College London\\
\email{nil.stolt@tum.de}}
\maketitle              
\begin{abstract}
Segmentation of anatomical shapes from medical images has taken an important role in the automation of clinical measurements. While typical deep-learning segmentation approaches are performed on discrete voxels, the underlying objects being analysed exist in a real-valued continuous space. Approaches that rely on convolutional neural networks (CNNs) are limited to grid-like inputs and not easily applicable to sparse or partial measurements. We propose a novel family of image segmentation models that tackle many of CNNs' shortcomings: Neural Implicit Segmentation Functions (NISF). Our framework takes inspiration from the field of neural implicit functions where a network learns a mapping from a real-valued coordinate-space to a shape representation. NISFs have the ability to segment anatomical shapes in high-dimensional continuous spaces. Training is not limited to voxelized grids, and covers applications with sparse and partial data. Interpolation between observations is learnt naturally in the training procedure and requires no post-processing. Furthermore, NISFs allow the leveraging of learnt shape priors to make predictions for regions outside of the original image plane. We go on to show the framework achieves dice scores of $0.87 \pm 0.045$ on a (3D+t) short-axis cardiac segmentation task using the UK Biobank dataset. We also provide a qualitative analysis on our frameworks ability to perform segmentation and image interpolation on unseen regions of an image volume at arbitrary resolutions. 

\end{abstract}
\section{Introduction}
Image segmentation is a core task in domains where the area, volume or surface of an object is of interest. The principle of segmentation involves assigning a class to every presented point in the input space. Typically, the input is presented in the form of images: aligned pixel (or voxel) grids, with the intention to obtain a class label for each. In this context, the application of deep learning to the medical imaging domain has shown great promise in recent years. With the advent of the U-Net \cite{unet}, Convolutional Neural Networks (CNN) have been successfully applied to a multitude of imaging domains and achieved (or even surpassed) human performance \cite{isensee2021nnu}. The convolution operation make CNNs an obvious choice for dealing with inputs in the form of 2D pixel- or 3D voxel-grids.

\begin{figure}
\includegraphics[width=\textwidth]{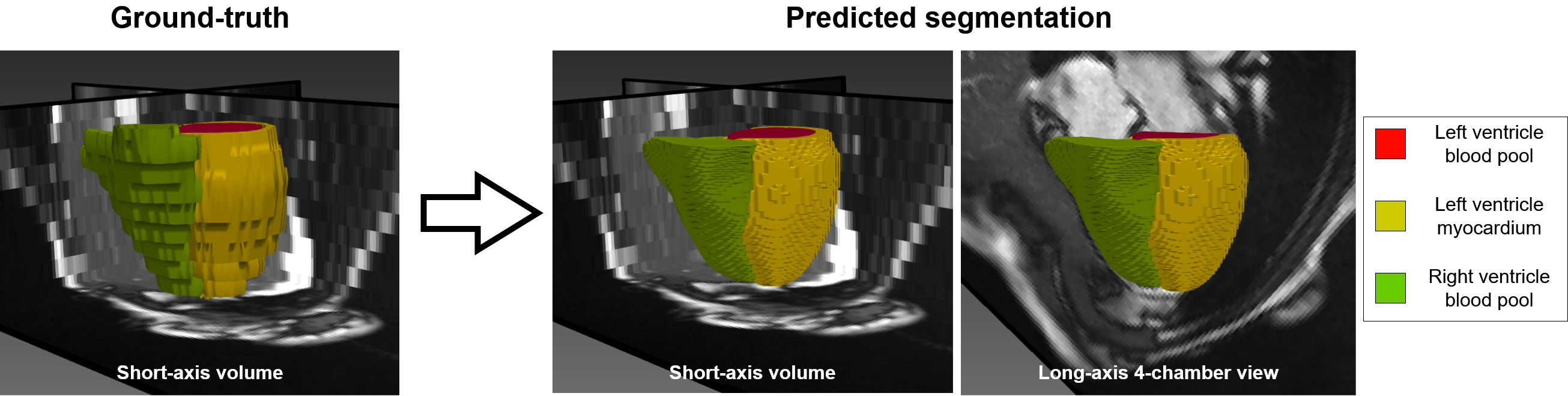}
\centering
\caption{Short axis volumes have low resolution along the ventricle's long axis. Given a short axis image volume, a NISF can produce arbitrary resolution segmentations along the long axis. } 
\label{SAX_diagram}
\end{figure}

Despite their efficacy, CNNs suffer from a range of limitations that lead to incompatibilities for some imaging domains. CNNs are restricted to data in the form of grids, and cannot easily handle sparse or partial inputs. Moreover, due to the CNN's segmentation output also being confined to a grid, obtaining smooth object surfaces requires post-processing heuristics. Predicting a high resolution segmentations also has implications on the memory and compute requirements in high-dimensional domains. Finally, the learning of long-distance spatial correlations requires deep stacks of layers, which may pose too taxing in low resource domains.


We introduce a novel approach to image segmentation that circumvents these shortcomings: Neural Implicit Segmentation Functions (NISF). Inspired by ongoing research in the field of neural implicit functions (NIF), a neural network is taught to learn a mapping from a coordinate space to any arbitrary real-valued space, such as segmentation, distance function, or image intensity. While CNNs employ the image's pixel or voxel intensities as an input, NISF's input is a real-valued vector $c\in\mathbb{R}^N$ for a single N-dimensional coordinate, alongside a subject-specific latent representation vector $h\in\mathbb{R}^d$. Given $c$ and $h$, the network is taught to predict image intensity and segmentation value pairs. The space $\mathcal{H}$ over all possible latent vectors $h$ serves as a learnable prior over all possible subject representations. 

In this paper, we describe an auto-decoder process by which a previously unseen subject's pairs of coordinate-image intensity values $(c, i)$ may be used to approximate that subject's latent representation $h$. Given a latent code, the intensity and segmentation predictions from any arbitrary coordinates in the volume may be sampled. We evaluate the proposed framework's segmentation scores and investigate its generalization properties on the UK-Biobank cardiac magnetic resonance imaging (MRI) short-axis dataset. We make the source code publicly available\footnote{Code repository: \url{https://github.com/NILOIDE/Implicit_segmentation}}.

\section{Related Work}

\noindent\textbf{Cardiac MRI.} \quad 
Cardiac magnetic resonance imaging (MRI) is often the preferred imaging modality for the assessment of function and structure of the cardiovascular system. This is in equal parts due to its non-invasive nature, and due to its high spatial and temporal resolution capabilities. The short-axis (SAX) view is a (3+t)-dimensional volume made up of stacked cross-sectional (2D+t) acquisitions which lay orthogonal to the ventricle's long axis (see Figure~\ref{SAX_diagram}). Spatial resolution is highest in-plane (typically $<$3mm$^2$), with a much lower inter-slice resolution (~10mm), and a temporal resolution of $\leq$45ms~\cite{kramer2020}. On the other hand, long-axis (LAX) views are (2D+t) acquisitions orthogonal to the SAX plane and provide high resolution along the ventricle's long axis.\\

\noindent\textbf{Image segmentation.} \quad
The capabilities of the CNN has caused it to become the predominant choice for image segmentation tasks~\cite{resnet,unet}. However, a pitfall of these models is their poor generalization to certain input transformations. One such transformation is scaling. This drawback limits the use of CNNs on domains with large variations in pixel spacings. Past works have attempted to mitigate this issue by accounting for dataset characteristics~\cite{isensee2021nnu}, building resilience through augmentations~\cite{zhao2019data}, or using multi-scale feature extractors~\cite{chen2016attention}.

Additionally, segmentation performed by fully convolutional model is restricted to predicting in pixel (or voxel) grids. This requires post-processing heuristics to extract smooth object surfaces. Works such as~\cite{Peng2020ECCV,khan2022} try to mitigate this issue through point-wise decoders that operate on interpolated convolutional features. Alternatives to binarized segmentation have been recently proposed such as soft segmentations~\cite{gros2021softseg} and distance field predictions~\cite{Dai2017,Stutz2018}. Smoothness can also be improved by predicting at higher resolutions. This is however limited by the exponential increase of memory that comes with high-dimensional data. Partitioning of the input can make memory requirements manageable~\cite{Bali2015,HouSKGDS15}, but doing so disallows the ability to learn long-distance spatial correlations.\\

\noindent\textbf{Neural implicit functions.} \quad 
In recent years, NIFs have achieved notable milestones in the field of shape representations \cite{NeRF,DeepSDF}. NIFs have multiple advantages over classical voxelized approaches that makes them remarkably interesting for applications in the medical imaging domain~\cite{huang2022neural,wolterink2022implicit}. First, NIFs can sample shapes at any points in space at arbitrary resolutions. This makes them particularly fit for working with sparse, partial, or non-uniform data. Implicit functions thus remove the need for traditional interpolation as high-resolution shapes are learnt implicitly by the network~\cite{amiranashvili2022}. This is specially relevant to the medical imaging community, where scans may have complex sampling strategies, have missing or unusable regions, or have highly anisotropic voxel sizes. These properties may further vary across scanners and acquisition protocols, making generalization across datasets a challenge. Additionally, the ability to process each point independently allows implicit functions to have flexible optimization strategies, making entire volumes be optimizable holistically.\\

\noindent\textbf{Image priors.} \quad 
The typical application of a NIF involves the training of a multi-layer perceptron (MLP) on a \emph{single} scene. Although generalization still occurs in generating novel views of the target scene, the introduction of prior knowledge and conditioning of the MLP is subject to ongoing research~\cite{amiranashvili2022,Klocek2019,Mehta2021,DeepSDF,MetaSDF,Siren}. Approaches such as~\cite{amiranashvili2022,DeepSDF} opt for auto-decoder architectures where the network is modulated by latent code at the input level. At inference time, the latent code of the target scene is optimized by backpropagation. Works such as~\cite{Mehta2021} choose to instead modulate the network at its activation functions. Other frameworks obtain the latent code in a single-shot fashion through the use of an encoder network~\cite{Klocek2019,Mehta2021,Siren,MetaSDF}. This latent code is then used by a hyper-network~\cite{Klocek2019,Mehta2021,Siren} or a meta-learning approach~\cite{MetaSDF} to generate the weights of a decoder network.

\section{Methods}

\noindent\textbf{Shared Prior.} \quad
In order to generalize to unseen subjects, we attempt to build a shared prior $\mathcal{H}$ over all subjects. This is done by conditioning the classifier with a latent vector $h \in \mathbb{R}^d$ at the input level. Each individual subject $j$ in a population $X$, can be thought of having a distinct $h_j$ that serves as a latent code of their unique features. Following \cite{amiranashvili2022,DeepSDF}, we initialize a matrix $H \in \mathbb{R}^{Xd}$, where each row is a latent vector $h_j$ corresponding to a single subject $j$ in the dataset. The latent vector $h_j$ of a subject is fed to the MLP alongside a point's coordinate and can be optimized through back-propagation. This allows $\mathcal{H}$ to be optimized to capture useful inter-patient features.\\

\noindent\textbf{Model Architecture.} \quad
The architecture is composed of a segmentation function $f_\theta$ and a reconstruction function $f_\phi$. At each continuous-valued coordinate $c~\in~\mathbb{R}^{N}$, function $f_\theta$ models the shape's segmentation probability $s_c$ for all $M$ classes, and function $f_\phi$ models the image intensity $i_c$. The functions are conditioned by a latent vector $h$ at the input level as follows:

\begin{equation}
    f_\theta: \left(c \in \mathbb{R}^{N}\right) \times \left(h \in \mathbb{R}^{d}\right) \rightarrow s_c \in [0, 1]^M, \quad \sum^{M}_{i=1} s_c^i = 1
\end{equation}

\begin{equation}
    f_\phi: \left(c \in \mathbb{R}^{N}\right) \times \left(h \in \mathbb{R}^{d}\right) \rightarrow i_c \in [0, 1]
\end{equation}

\begin{figure}[h!]
\makebox[\textwidth][c]{\includegraphics[width=1.\textwidth]{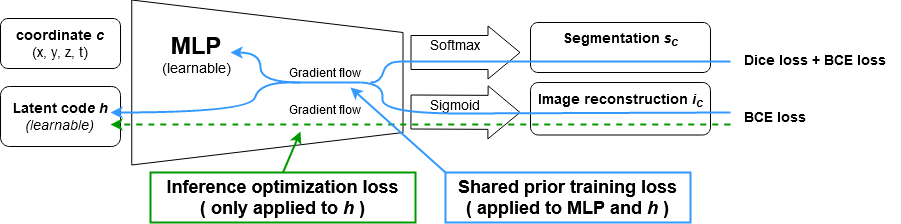}}%
\centering
\caption{Training and inference setups. During the prior's training, the MLP and the input latent code $h$ are jointly optimized on image reconstruction and segmentation losses (solid blue line). At inference, solely the latent code $h$ is optimized exclusively on the image reconstruction (dashed green line).} 
\label{Architecture}
\end{figure}

In order to improve local agreement between the segmentation and reconstruction functions, we jointly model $f_\theta$ and $f_\phi$ by a unique multi-layer perceptron (MLP) with two output heads (Figure~\ref{Architecture}). We employ Gabor wavelet activation functions~\cite{WIRE} which are known to be more expressive than Fourier Features combined with ReLU~\cite{tancik2020fourier} or sinusoidal activation functions \cite{Siren}. \\

\noindent\textbf{Prior Training.} \quad
Following the setup described in~\cite{amiranashvili2022}, we randomly initialize the matrix $H$ consisting of a trainable latent vector $h_j \sim \mathcal{N}\left(0,10^{-2}\right)$ for each subject in the training set. On each training sample, the parameters of the MLP are jointly optimized with the subject's $h_j$. We select a training batch by uniformly sampling a time frame $t$ and using all points within that 3D volume. Each voxel in the sample is processed in parallel along the batch dimension. Coordinates are normalized to the range $[0,1]$ based on the voxel's relative position. 

The difference in image reconstruction from the ground-truth voxel intensities is supervised using binary cross-entropy (BCE). This is motivated by our data's voxel intensity distribution being heavily skewed towards the extremes. The segmentation loss is a sum of a BCE loss component and a Dice loss component. We found that adding a weighting factor of $\alpha = 10$ to the image reconstruction loss component yielded inference-time improvements on both image reconstruction and segmentation metrics. Additionally, L2 regularization is applied to the latent vector $h_j$ and the MLP's parameters. The full loss is summarized as follows:
\begin{equation}
\begin{aligned}
    \mathcal{L}_{train}(\theta,\phi,h_j) \;=\; \mathcal{L}_{BCE}\Big(f_\theta(c, h_j), s_c\Big) \;+\; \mathcal{L}_{Dice}\Big(f_\theta(c, h_j), s_c\Big)\qquad\qquad \\
    \qquad\qquad+\;\; \alpha \: \mathcal{L}_{BCE}\Big(f_\phi(c, h_j), i_c\Big) \;+\; \mathcal{L}_{L2}(\theta) \;+\; \mathcal{L}_{L2}(\phi) \;+\; \mathcal{L}_{L2}(h_j)
\end{aligned}
\end{equation}

\noindent\textbf{Inference.} \quad 
Once the segmentation function $f_\theta$ has learnt a mapping from the population prior $\mathcal{H}$ to the segmentation space $S$, inference becomes a task of finding a latent code $h$ within $\mathcal{H}$ that correctly models the new subject's features. The ground-truth segmentation of a new subject is obviously not available at inference, and it is thus not possible to use $f_\theta$ to optimize $h$. However, since both functions $f_\phi$ (image reconstruction) and $f_\theta$ (segmentation) have been jointly trained by consistently using the same latent vector $h$, we make the following assumption: \textit{A latent code $h$ optimized for image reconstruction under $f_\phi$ will also produce accurate segmentations under $f_\theta$}. This assumption makes it possible to use the image reconstruction function $f_\phi$ alone to find a latent code $h$ for an unseen image in order to generalize segmentation predictions using $f_\theta$.

For this task, a new $h \sim \mathcal{N}\left(0,10^{-4}\right)$ is initialized. The weights of the MLP are frozen, such that the only tuneable parameters are those of $h$. Optimization is performed exclusively on the image reconstruction loss (dashed green line in Figure~\ref{Architecture}): 
\begin{equation}
\begin{aligned}
    \mathcal{L}_{infer}(h_j) \;=\; \mathcal{L}_{BCE}\Big(f_\phi(c, h_j), i_c\Big) \;+\; \mathcal{L}_{L2}(h_j)
\end{aligned}
\end{equation}

Due to the loss being composed exclusively by the image reconstruction term, $h$ is expected to eventually overfit to $f_\phi$. Special care should be taken to find a step-number hyperparameter that stops the optimization of $h$ at the optimal segmentation performance. In our experiments, we chose this parameter based on the Dice score of the best validation run.

\section{Experiments and Results}

\noindent\textbf{Data overview.} \quad
The dataset consists of a random subset of 1150 subjects from the UK Biobank's short-axis cardiac MRI acquisitions~\cite{UKBB}. An overview of the UK Biobank cohort's baseline statistics can be found in their showcase website~\cite{UKBBshowcase}. The dataset split included 1000 subjects for the prior training, 50 for validation, and 100 for testing. The (3D+t) short-axis volumes are anisotropic in nature and have a wide range of shapes and pixel spacings along the spatial dimensions. No form of preprocessing was performed on the images except for an intensity normalization to the range $[0,1]$ as performed in similar literature~\cite{Bai2018}. The high dimensionality of (3D+t) volumes makes manual annotation prohibitively time consuming. Due to this, we make use of synthetic segmentation as ground truth shapes created using a trained state of the art segmentation CNN provided by~\cite{Bai2018}. The object of interest in each scan is composed of three distinct, mutually exclusive sub-regions: The left ventricle (LV) blood pool, LV myocardium, and right ventricle (RV) blood pool (see Figure \ref{SAX_diagram}).\\

\noindent\textbf{Implementation details.} \quad 
The architecture consists of 8 residual layers, each with 128 hidden units. The subject latent codes had 128 learnable parameters. The model was implemented using Pytorch and trained on an NVIDIA A40 GPU for 1000 epochs, lasting approximately 9 days. Inference optimization lasted 3-7 minutes per subject depending on volume dimensions. Losses are minimized using the ADAM optimizer~\cite{ADAM} using a learning rate of $10^{-4}$ during the prior training training and $10^{-4}$ during inference. \\

\begin{figure}[h!]
\makebox[\textwidth][c]{\includegraphics[width=.95\textwidth]{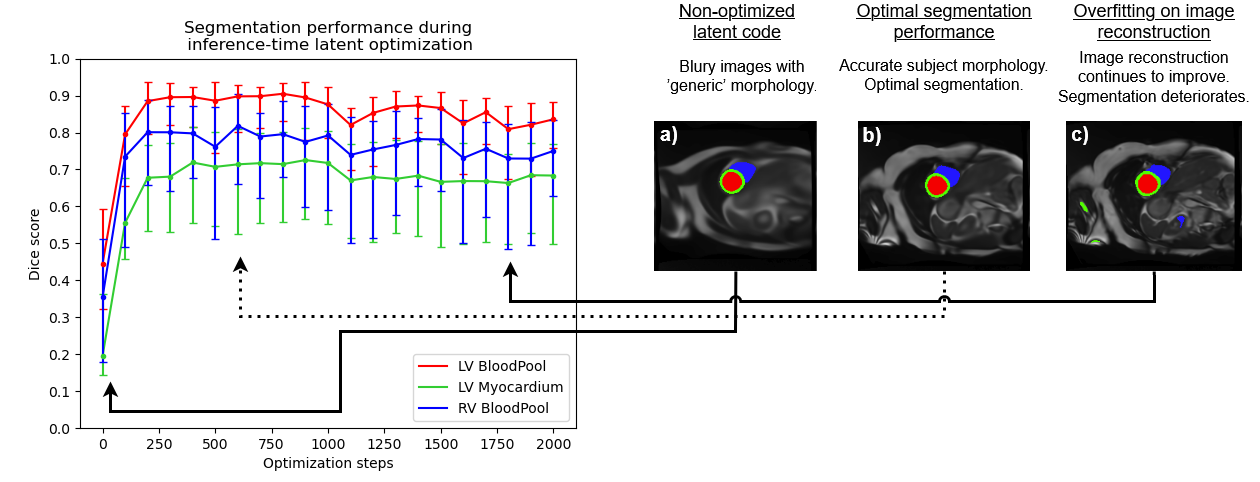}}%
\centering
\caption{Segmentation Dice trend during a subject's inference. Early stopping is important to prevent overfitting on reconstruction task. a) Non-optimized latent code creates blurry images with 'generic' morphology. b) As the latent code is optimized, subject morphology begins to be accurately reconstructed. Segmentation performance reaches an optimum. c) Reconstruction continues to improve, but segmentation deteriorates.} 
\label{Figure:Inference_scores}
\end{figure}

\noindent\textbf{Results.} \quad
As the latent code is optimized during inference, segmentation metrics follow an overfitting pattern (see Figure~\ref{Figure:Inference_scores}). This is an expected consequence of the inference process optimizing solely on the image reconstruction loss. Early stopping should be employed to obtain the best performing latent code state.

\begin{figure}[h!]
\makebox[\textwidth][c]{\includegraphics[width=.90\textwidth]{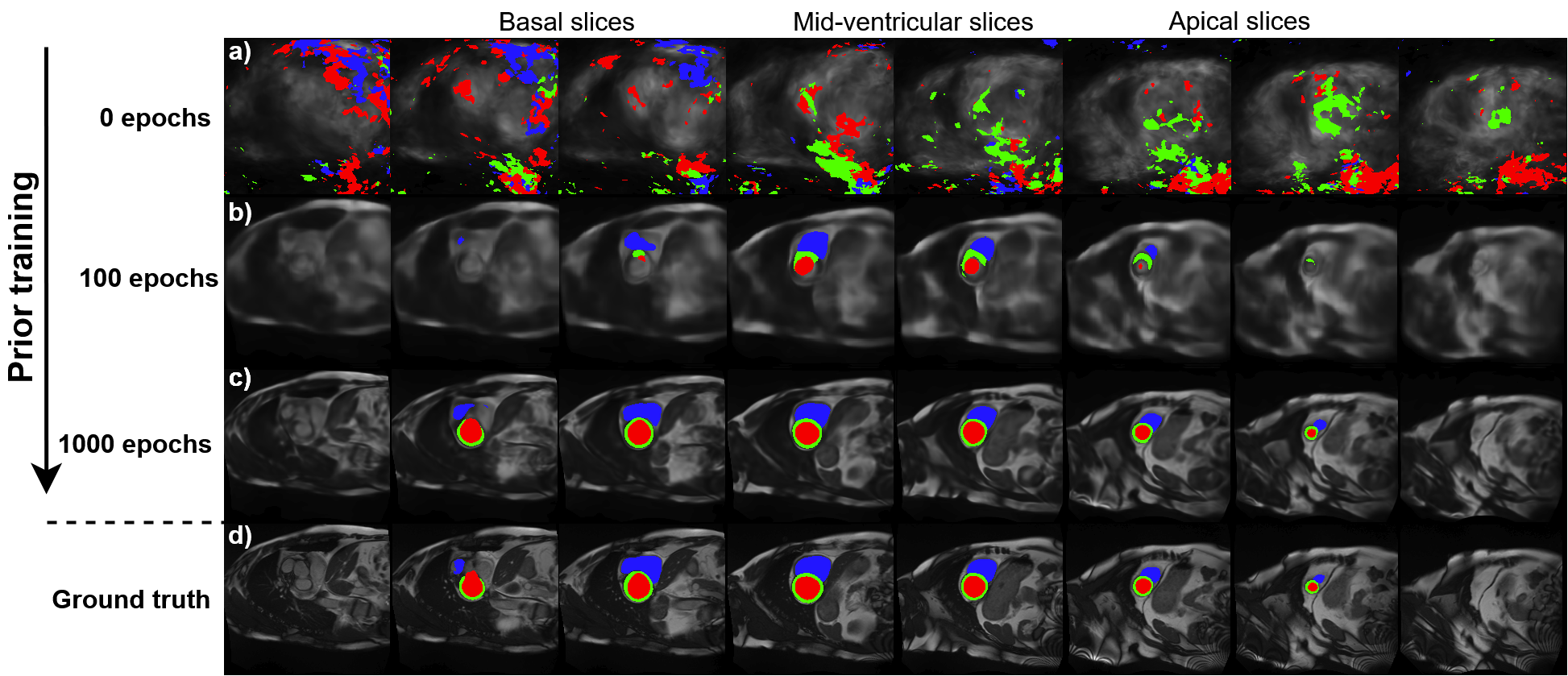}}%
\centering
\caption{Inference-time segmentation and image reconstruction at various stages of the prior's training process. a) Prior has not been trained. Inference can roughly reconstruct the image outline. Segmentation fails. b) Early on, reconstructed images are blurry. Segmentation is poor, but at the correct region. c) Eventually images are reconstructed with great detail and segmentations are accurate. d) Ground truth.} 
\label{Figure:Predictions}
\end{figure}

The benefits of training a prior over the population is investigated by tracking inference-time Dice scores obtained from spaced-out validation runs. Training of the prior is shown to significantly improve performance of segmentation and image reconstruction at inference-time as seen in Figure~\ref{Figure:Predictions}.

\begin{table}[h!]
\caption{Class Dice scores for the 100 subject test dataset.}
\label{Table:results}
\setlength{\tabcolsep}{0.5em}
\begin{tabular}{|p{1.5cm}||p{2.2cm}|p{2.2cm}|p{2.3cm}|p{2.2cm}| }
 \hline
 \small{Class} & \small{Classes average} & \small{LV blood Pool} & \small{LV myocardium} & \small{RV blood Pool}\\
 \hline
 \small{Dice score} & \small{$0.87 \pm 0.045$} & \small{$0.90\pm 0.037$} & \small{$0.82\pm 0.075$} & \small{$0.88\pm 0.063$}\\
 \hline
\end{tabular}
\end{table}

Validation results showed the average optimal number of latent code optmimization steps at inference to be 672. Thus, the test set per-class Dice scores (Table~\ref{Table:results}) were obtained after 672 optimization steps on $h$ for each test subject. 

Further investigation is performed on the generalization capabilities of the subject prior by producing segmentations for held-out sections of the image volume. First, the subject's latent code is optimized using the inference process. Then, the model's output is sampled at the held-out region's coordinates.

\begin{figure}[h!]
\makebox[\textwidth][c]{\includegraphics[width=.95\textwidth]{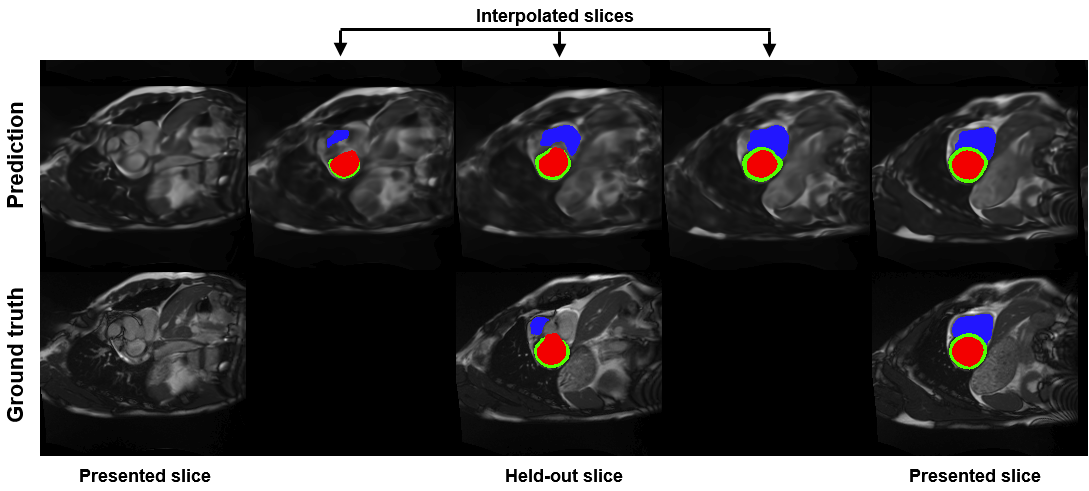}}%
\centering
\caption{Interpolation predictions for a held-out basal slice. Top row: Predicted segmentation overlayed on predicted image. Bottom row: Ground truth segmentation overlayed on original image. Middle column is never shown to network during inference. Black slices don't exist in original image volume. The model appears to understand how the ventricles come into view as we descend down the slice dimension.} 
\label{Figure:Segmentations_holdout}
\end{figure}

Right ventricle segmentation in basal slices is notoriously challenging to manually annotate due to the delineation of the atrial and ventricular cavity combined with the sparsity of the resolution along the long axis~\cite{Budai2020}. Nonetheless, as seen in Figure~\ref{Figure:Segmentations_holdout}, our approach is capable of capturing smooth and plausible morphology of these regions despite not having access to the image information.

\begin{figure}[h!]
\makebox[\textwidth][c]{\includegraphics[width=1.0\textwidth]{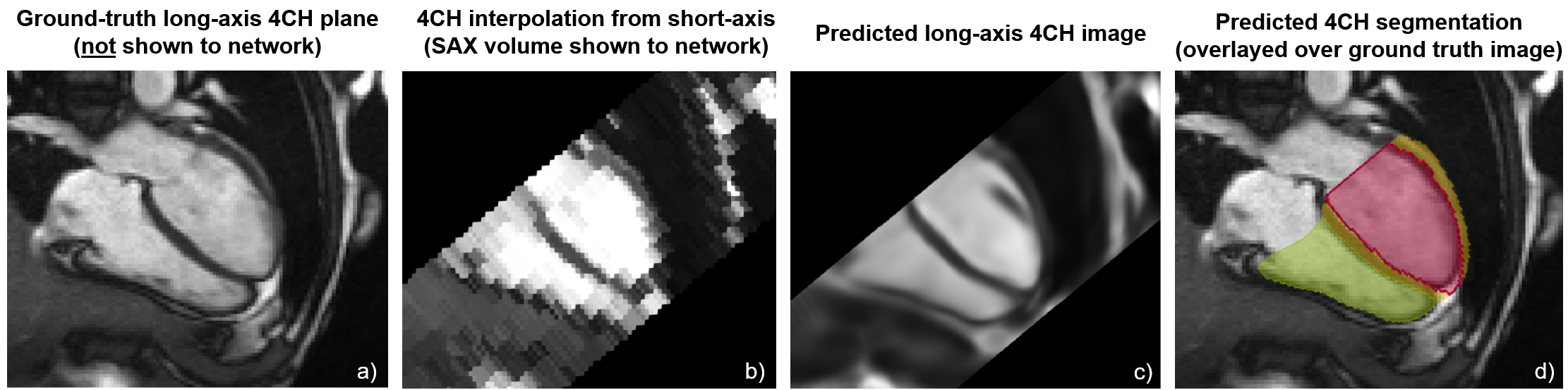}}%
\centering
\caption{Segmentation of a held-out long-axis 4-chamber plane from SAX image data. a)~Ground-truth long-axis 4-chamber view (\underline{not} presented to network). b)~Nearest-neighbour interpolation of 4-chamber view from SAX volume. c)~Predicted 4-chamber image plane. d)~Predicted 4-chamber view segmentation.} 
\label{Figure:4CH_interpolation}
\end{figure}

We go on to show NISF's ability to generate high-resolution segmentation for out-of-plane views. We optimize on a short-axis volume at inference and subsequently sample coordinates corresponding to long-axis views. Despite \textit{never} presenting a ground-truth long-axis image, the model reconstructs an interpolated view and provides an accurate segmentation along its plane (Figure~\ref{Figure:4CH_interpolation}).

\section{Conclusion}
We present a novel family of image segmentation models that can model shapes at arbitrary resolutions. The approach is able to leverage priors to make predictions for regions not present in the original image data. Working directly on the coordinate space has the benefit of accepting high-dimensional sparse data, as well as not being affected by variations in image shapes and resolutions. We implement a simple version of this framework and evaluate it on a short-axis cardiac MRI segmentation task using the UK Biobank. Reported Dice scores on 100 unseen subjects average $0.87 \pm 0.045$. We also perform a qualitative analysis on the framework's ability to predict held-out sections of image volumes.

\section{Acknowledgements}
This work is funded by the Munich Center for Machine Learning and European Research Council (ERC) project Deep4MI (884622). This research has been conducted using the UK Biobank Resource under Application Number 87802.


%
%
%
\bibliographystyle{splncs04}
\bibliography{bibliography}
\end{document}